\documentclass[sigconf]{acmart}
\AtBeginDocument{%
  \providecommand\BibTeX{{%
    \normalfont B\kern-0.5em{\scshape i\kern-0.25em b}\kern-0.8em\TeX}}}

\begin{document}

\title[A Comparison of Immersive Analytics with Augmented and Virtual Reality]{I Did Not Notice: A Comparison of Immersive Analytics with Augmented and Virtual Reality}


\author{Xiaoyan Zhou}
\affiliation{%
  \institution{Colorado State University}
  \city{Fort Collins}
  \state{CO}
  \country{USA}}
\email{Xiaoyan.Zhou@colostate.edu}

\author{Anil Ufuk Batmaz}
\affiliation{%
  \institution{Concordia University}
  \city{Montreal}
  \state{QC}
  \country{Montreal}}
\email{ufuk.batmaz@concordia.ca}

\author{Adam S. Williams}
\affiliation{%
  \institution{Colorado State University}
  \city{Fort Collins}
  \state{CO}
  \country{USA}}
\email{AdamWil@Colostate.edu}

\author{Dylan Schreiber}
\affiliation{%
  \institution{Colorado State University}
  \city{Fort Collins}
  \state{CO}
  \country{USA}}
\email{dylan.schreiber@colostate.edu}

\author{Francisco Ortega}
\affiliation{%
  \institution{Colorado State University}
  \city{Fort Collins}
  \state{CO}
  \country{USA}}
\email{fortega@colostate.edu}

\renewcommand{\shortauthors}{Trovato and Tobin, et al.}

\begin{abstract}
  Immersive environments enable users to engage in embodied interaction, enhancing the sensemaking processes involved in completing tasks such as immersive analytics. Previous comparative studies on immersive analytics using augmented and virtual realities have revealed that users employ different strategies for data interpretation and text-based analytics depending on the environment. Our study seeks to investigate how augmented and virtual reality influences sensemaking processes in quantitative immersive analytics. Our results, derived from a diverse group of participants, indicate that users demonstrate comparable performance in both environments. However, it was observed that users exhibit a higher tolerance for cognitive load in VR and travel further in AR. Based on our findings, we recommend providing users with the option to switch between AR and VR, thereby enabling them to select an environment that aligns with their preferences and task requirements.
\end{abstract}



\begin{teaserfigure}
  \includegraphics[width=\textwidth]{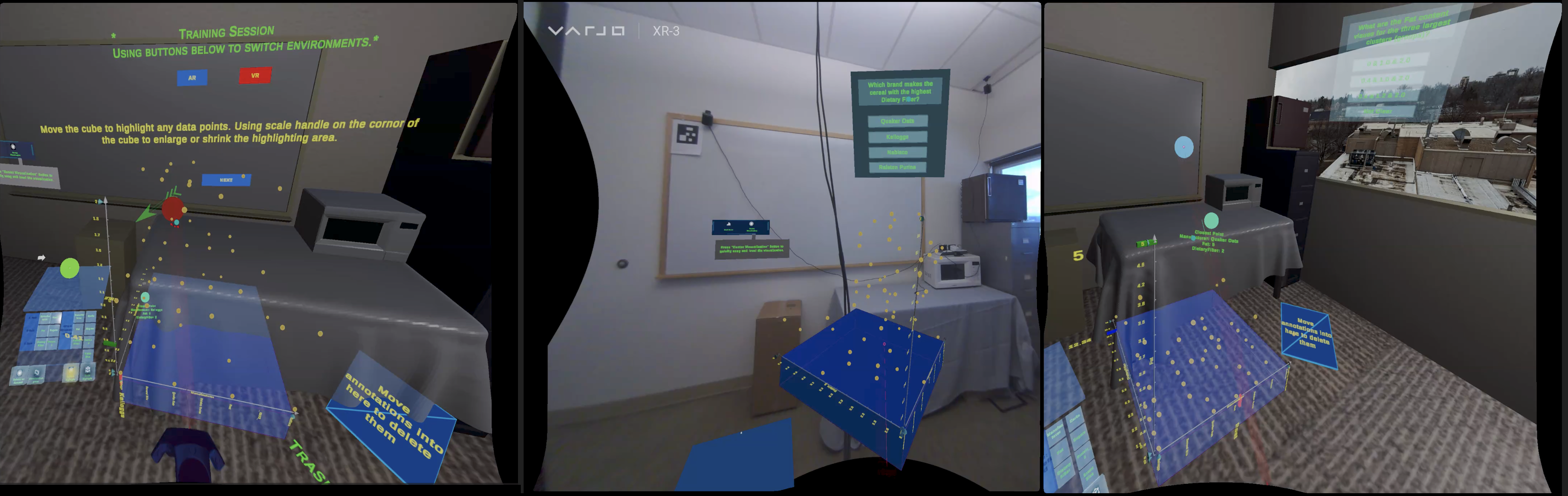}
  \caption{The interface displayed in the training session (left), AR session (middle), and VR session (right)}
  \Description{The interface shown in the training session, AR session, and VR session}
  \label{fig:teaser}
\end{teaserfigure}


\maketitle

\section{Introduction}
Along with the rapid growth in the popularity of virtual reality  (VR) and mixed reality (MR) headsets, many researchers are particularly interested in exploring the differences between augmented reality (AR) and VR Head Mounted Displays (HMDs) and how they can be leveraged to enhance user performance~\cite{STE+19, LIS+23, WHI+20, WIL+23}. While see-through AR provides users with an interactive and intuitive experience by seeing 3D objects and their own hands, VR offers users a fully immersive experience with clutter-free and infinite spaces. Previous studies have explored the effects of AR or VR environments on sensemaking strategies in non-quantitative immersive analytics~\cite{LIS+23} and their impact on the perception of 3D data visualization ~\cite{WHI+20}. However, there is limited empirical evidence on how different realities influence 3D data visualization sensemaking, as well as their respective advantages and disadvantages.
To address this gap, we developed a data analytics platform in both AR and VR with the same HMD, using identical objects and layouts. We then conducted a user study to measure performance and record feedback in each reality. To minimize confounding variables such as differing fields of view, we utilized a headset that supports displays for both AR and VR.
Our results reveal that user performance, overall subjective workload, and presence experience are comparable between AR and VR. However, users reported distinct perceptions of frustration in AR versus VR; additionally, they displayed more varied postures and physical navigation in the VR environment. Based on our observations, we offer insights and guidelines for future researchers developing immersive analytics platforms to consider.

\section{Related Work}
\subsection{Immersive Analysis}
Immersive analytics extends the domains of data visualization, visual analytics, mixed reality, computer graphics, and human-computer interaction. Users engage in sensemaking within immersive environments using abstract data representations to facilitate a comprehensive understanding of data and decision-making processes for independent work as well as collaborative endeavors ~\cite{MAR+18, CHA+15, SKA+19}. 

Recent studies have explored user performance in immersive analytics with different realities ~\cite{BAC+18, BAT+19} while also focusing on improving sense-making strategies to increase user performance and experience when engaging in immersive analysis tasks~\cite{COR+17, SAT+20, ZHA+23, TAH+22}. For example, several studies have shown that the use of spherical layouts for displaying multiple visualizations can enhance the sensemaking process and encourage greater user engagement ~\cite{BAT+19, SAT+20, LIU+20, LIS+21}.

Although earlier work primarily focused on sensemaking with quantitative datasets ~\cite{COR+17, BAT+19}, researchers suggested that non-quantitative data could also benefit from the significant space available in immersive environments for offloading cognition during a sensemaking process~\cite{AND+10, YAN+22}. Our research focuses on immersive data visualization with a 3D scatter plot to explore variations in users' sensemaking strategies in AR and VR settings.

\subsection{AR vs. VR}
Several previous studies have compared AR and VR focused on specific aspects such as object manipulations~\cite{KRI+18, batmaz2022effect}, data visualization interpreting ~\cite{WHI+20}, depth estimations ~\cite{PIN+19}, eye-hand coordination ~\cite{BAT+20}, mode switching between display types ~\cite{SER+22}, or different research methods ~\cite{VOL+19}. Steffen et al. provided guidelines for developing specific affordances for better user experiences by comparing the affordances between AR, VR, and physical reality~\cite{STE+19}. Williams et al. initiated discussions about how different levels of reality choices may impact user interactions while performing immersive analytics~\cite{WIL+23}.

The most related study to this one came from Lisle et al. ~\cite{LIS+23}, who explored the sensemaking task in an immersive analytics context with AR and VR. They found that users can focus more on the task in VR than AR; however, with the ability to access real-world tools, user experience has improved during the task execution. Differently, our study focuses on immersive data visualization instead of text-based data. This could lead to different approaches to user navigation and sensemaking processes due to the spread of the data. Davidson et al. found that navigation patterns change according to the different stages of the sensemaking process ~\cite{DAV+22}. 

Cross-virtuality analytics has become increasingly prominent in recent years, with the goal of enabling users to transition seamlessly between different levels of mixed reality. This approach aims to address the limitations of each technology and facilitate a combination of their advantages for enhanced analysis efficiency ~\cite{RIE+20, FRO+22}. Our findings indicate that the choice of reality level depends on individual preferences and the complexity of the analysis task at hand. We offer insights and guidelines for future researchers looking to implement immersive analytics in cross-virtuality environments.

\section{Study Design}
We employed a within-design approach with two display types ($DT_2=$VR, AR). Each participant completed two question sets ($QS_2=$ VR questions, AR Questions), 6 questions in AR and 6 in VR, with the same data set. We ensured an even distribution of questions across both conditions concerning question type and difficulty level. The order of display type plus question set was counterbalanced among participants to mitigate learning effects, as was the order of questions for each participant under the same condition. Besides objective metrics such as task completion time and accuracy of answers, we used subjective measures including VZ-2 paper folding test~\cite{ekstrom1976manual}, NASA Task Load Index scale (NASA TLX)~\cite{HART1988139}, Presence Questionnaire (PQ)~\cite{presence}, and System Usability Scale (SUS)~\cite{brooke1996sus}. Additionally, a semi-interview was conducted at the end of each experiment to gather more insights into user experience.

\subsection{Experiment Setup}
The study utilized a Varjo XR-3 headset~\cite{Varjo2023}, which is the best mixed reality headset that is available on the market. This headset operated on a desktop PC equipped with an Intel Core i9-11900F processor and an NVIDIA GeForce RTX 3090 graphics card. Additionally, participants used HTC Vive controllers~\cite{VIVEController} to manipulate virtual objects in both environments, with tracking performed by three HTC Vive Tracker 2.0 units~\cite{tracker}. Our immersive analytics platform was employed in both conditions, and a simulated office environment was presented in VR, as seen in ~\autoref{fig:teaser}. To elicit natural interactions from participants, we included an office desk, corner chairs, a bookshelf, a microwave, and a mini refrigerator to replicate the real office setting. Furthermore, participants were provided with bar stools for easy body rotation as well as standing up or sitting down without safety concerns. 

\subsection{Procedure}
Upon the participant's arrival, the experimenter provided a brief introduction and handed consent forms for signatures. Then, participants completed a demographic questionnaire and the VZ-2 paper folding test. Next, the experimenter introduced the study and experiment process before instructing participants to put on headsets and use controllers to interact with virtual objects. They then followed the instructions displayed to complete eye calibration, training sessions, as well as two experimental sessions. During the training session, the participants were guided through each function of the platform with detailed instructions shown on the HMD. During the experimental session, participants were asked to answer 6 questions in each environment with the same dataset. These questions included outlier detection requiring axis switching or identifying data at a fourth dimension, identification of summarization (mean or median) and data trend, along with cluster recognition. Following each experimental session, participants filled out NASA TLX, PQ, and SUS. A 3-minute break was implemented between conditions. Each experiment finished with a semi-structured interview. The entire study takes around 60 minutes.

\subsection{Participants}
\begin{figure}
  \begin{minipage}[t]{0.8\linewidth}
    \centering
    \includegraphics[width=1\textwidth]{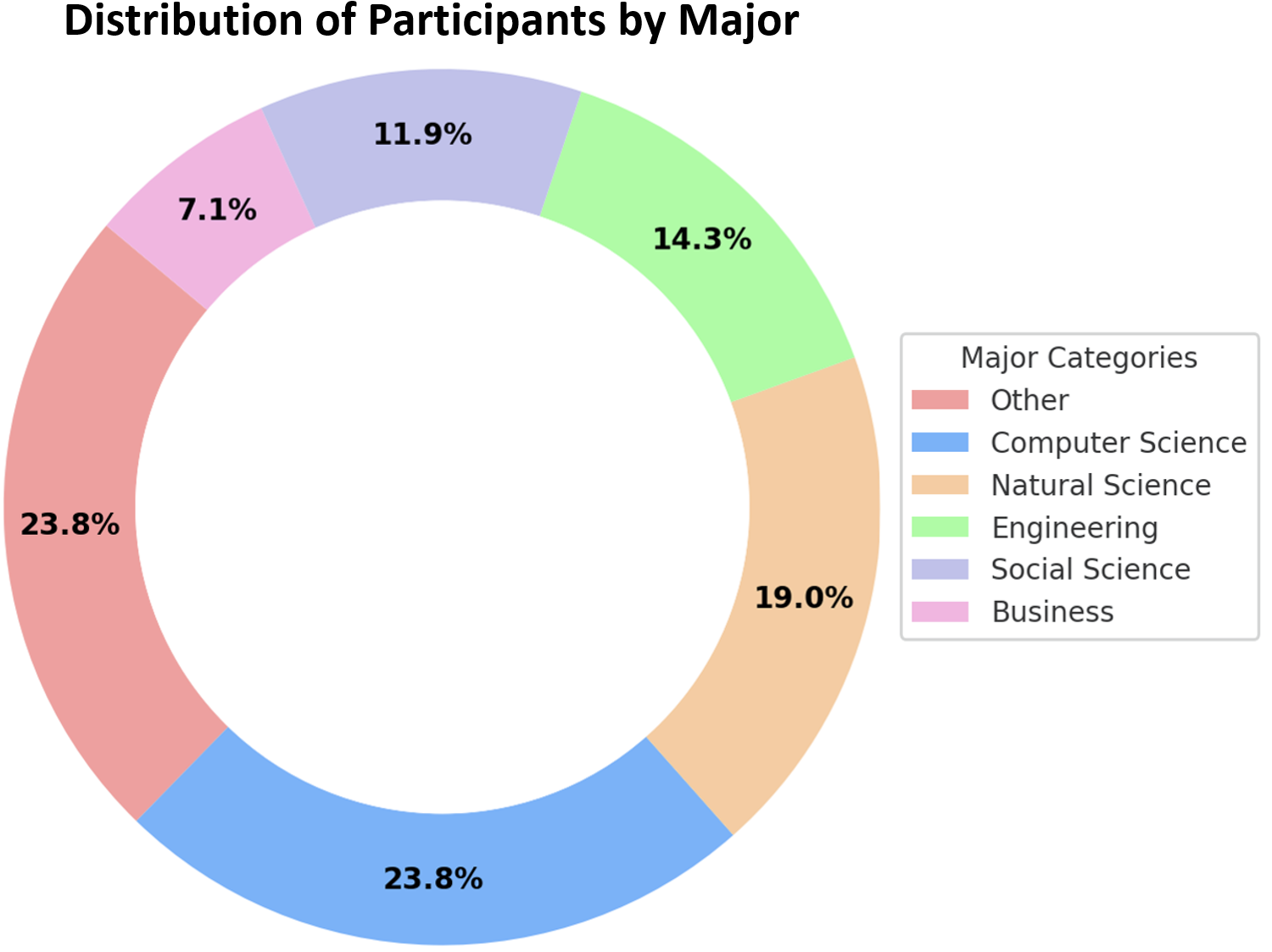}
    \caption{Participants Background in the Study}
    \Description{A donuts chart showing all majors that participants came from}
    \label{fig:particpants}
  \end{minipage}
  \end{figure}

40 participants (17 female, 20 male, 3 preferred not to specify their gender or identify as nonbinary) aged between 18 and 67 (average age of 29.45) volunteered for this experiment. Of the participants, 36 were right-handed while the rest were left-handed. Our participant pool comprised individuals from various fields: computer science, engineering, natural science, social science, business school, and others who were employees or did not specify a field of study (Figure~\ref{fig:particpants}). 

\subsection{Hypotheses}
Based on the results from the preliminary study~\cite{WIL+23A} and prior research comparing AR and VR~\cite{WHI+20, LIS+23}, we have developed three hypotheses for this study.

\textbf{H1:} Participants’ performance in AR is higher than in VR. Our preliminary study showed that participants in VR performed better than those in AR. Apart from variables like different head displays, we suspect that the presence of the experimenter and a cluttered room may influence performance, as suggested by a previous study~\cite{KRI+18, LIS+23}. Therefore, we eliminated unnecessary objects from the room and introduced a folding screen between the participants and the experimenter to see if these adjustments led to similar performances across both conditions. 

\textbf{H2:} Participants spend more time in AR than in VR. The preliminary study's results indicated that some participants experienced frustration during the experiment. It could have been caused by the difficulty of using hand gestures of HoloLens 2, as noted in another comparison work between AR and VR head-mounted displays ~\cite{GRE+18}. To address potential issues with hand gesture interactions, we provide controllers in both environments. Additionally, the visibility of the real world is anticipated to influence color perceptions among participants, possibly affecting their engagement with tasks in AR ~\cite{WHI+20, GUO+22}. It is also assumed that due to real-world distractions, participants will take longer to complete tasks in AR compared to VR.

\textbf{H3:} Participants in AR navigate through visualization physically more than VR. In the preliminary study, participants were given a chair equipped with a handle and asked to sit at a table before the experiment started. The results showed that only some VR participants exhibited upper body movement. Building on prior research, which suggests that with full environmental awareness, participants feel more comfortable moving around in AR than in VR~\cite{WHI+20}. By informing participants about the freedom of movement and implementing a bar stool setting in this study, we anticipated that greater physical movement in AR than in VR.

\section{Results}
After examining the correctness of each question in both conditions, we found that question 4 in the AR condition has a significantly lower correctness rate compared to other questions. It was also identified by the Interquartile Range method as an outlier, so we decided to remove question 4 from our dataset before proceeding further analysis. We performed a chi-square test to compare the proportion of correct responses between AR and VR conditions. It yielded a p-value of 0.405, suggesting no significant difference in correctness between the two display conditions. When comparing the time spent on questions between the two conditions, the Mann-Whitney U test indicated no significant difference in the time spent on answering questions across these two conditions (p-value = 0.745). These results indicate that hypothesis H1 is not supported. Moreover, there is no evidence supporting H2, which suggests participants spent more time in AR. 

\begin{figure}
  \begin{minipage}[t]{1\linewidth}
    \centering
    \includegraphics[width=1\textwidth]{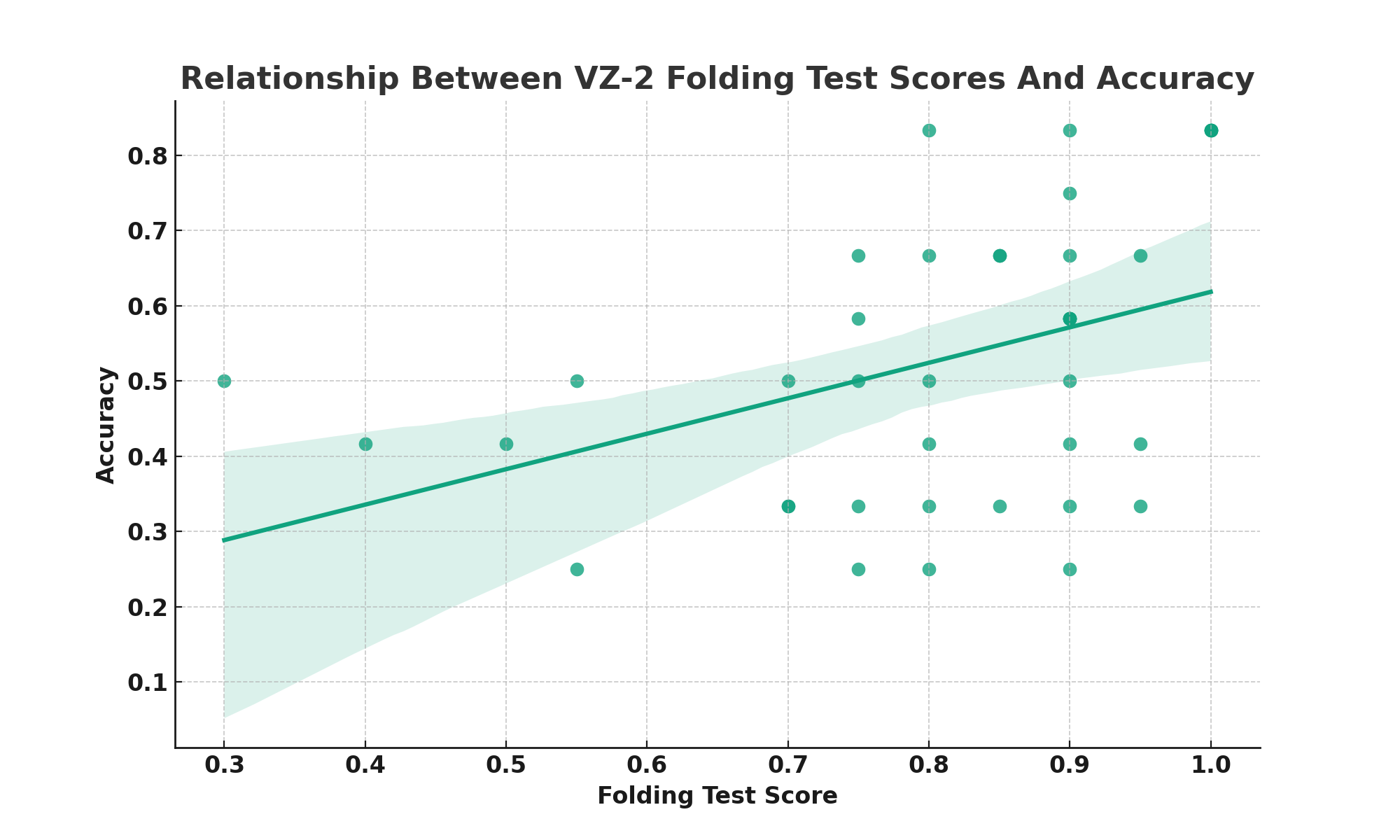}
    \caption{Regression Line of Folding Test and Accuracy}
    \Description{A regression line shows the relationship between folding test results and the accuracy of the answers in the experiment.}
    \label{fig:foldingtest}
  \end{minipage}
  \end{figure}

We conducted a correlation analysis to investigate the connection between scores from the VZ-2 paper folding test and the answers' accuracy in the experiment. The analysis uncovered a moderate positive correlation, with a correlation coefficient of 0.40, suggesting that participants who achieved higher scores on the folding test tended to have greater accuracy. However, this relationship was not strongly pronounced. After removing an outlier, further regression analysis supported these findings by showing a moderate positive linear relationship (~\autoref{fig:foldingtest}).
  \begin{figure}
  \begin{minipage}[t]{1\linewidth}
    \centering
    \includegraphics[width=1\textwidth]{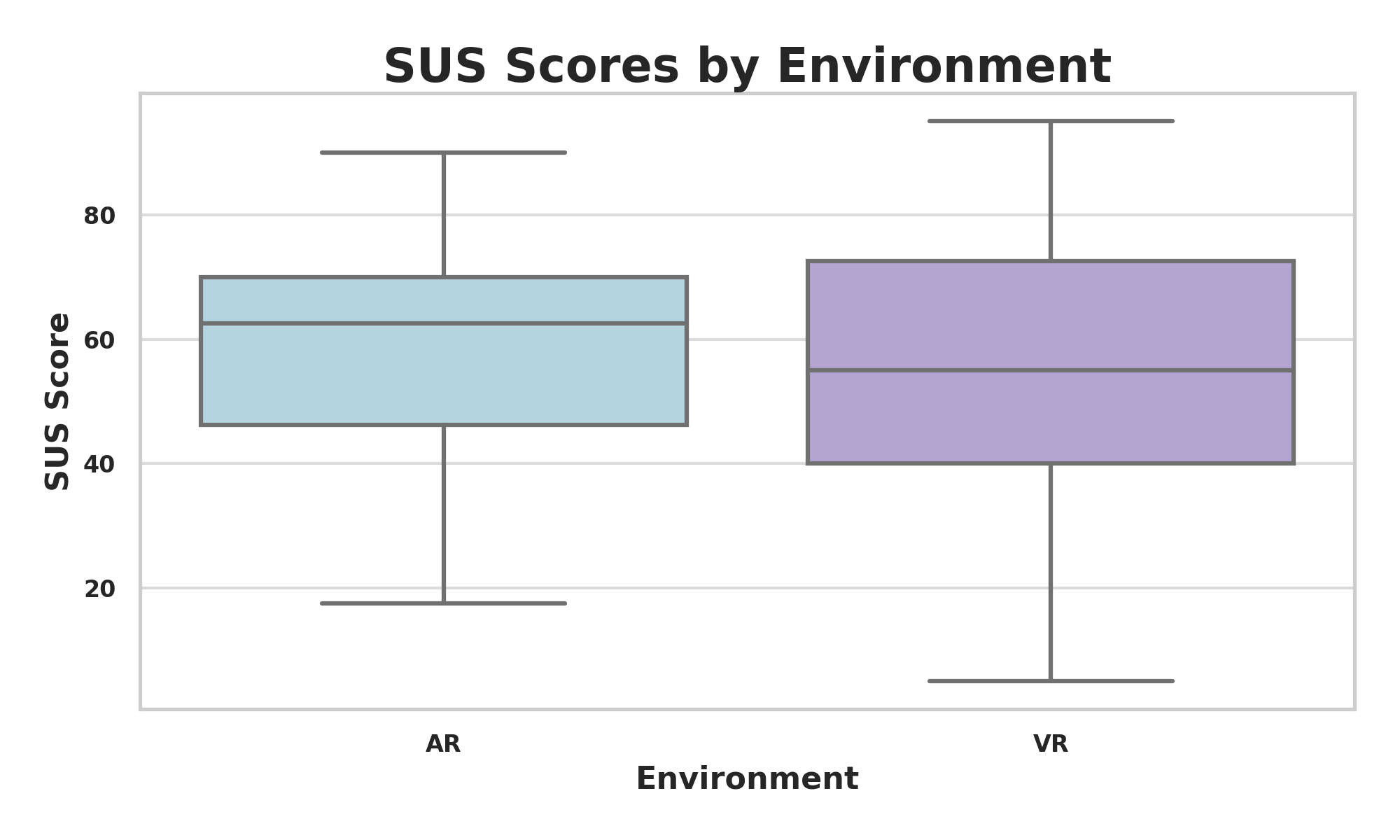}
    \caption{SUS Scores of AR and VR}
    \Description{A boxplot shows the sus score of AR and VR environments.}
    \label{fig:sus}
  \end{minipage}
  \end{figure}

A non-parametric Wilcoxon signed-rank test was conducted to compare the mean SUS scores between the AR and VR environments after removing one outlier in the AR condition. The non-significant p-value of 0.119 suggests that there is no statistically significant difference in system usability across the AR and VR environments. We attribute the comparatively low SUS scores (~\autoref{fig:sus}) to technical difficulties we encountered multiple times with XR-3 during the study, which were promptly resolved upon identification. Participants reported experiencing visual lagging and latency response from the headset once connection issues arose, likely impacting usability ratings. Additionally, a wide spread of SUS scores supports these technical challenges, as some users found the system highly usable while others experienced abnormal usability issues.

  \begin{figure}
  \begin{minipage}[t]{1\linewidth}
    \centering
    \includegraphics[width=1\textwidth]{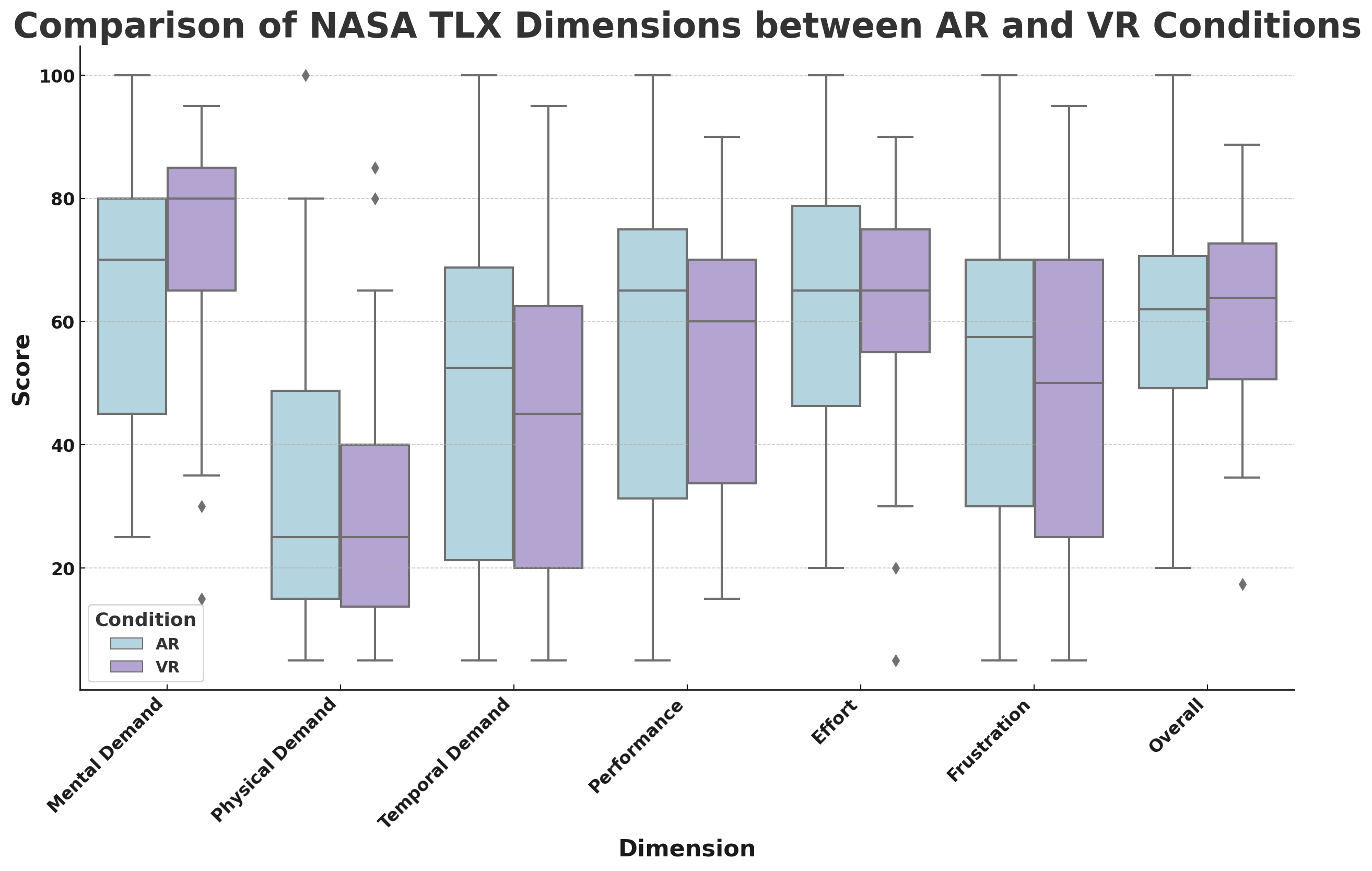}
    \caption{Comparison of NASA TLX Overall Scores}
    \Description{A boxplot shows that overall scores in AR and VR environments.}
    \label{fig:nasa_tlx_boxplot}
  \end{minipage}
\end{figure}

\begin{figure}
  \begin{minipage}[t]{1\linewidth}
    \centering
    \includegraphics[width=1\textwidth]{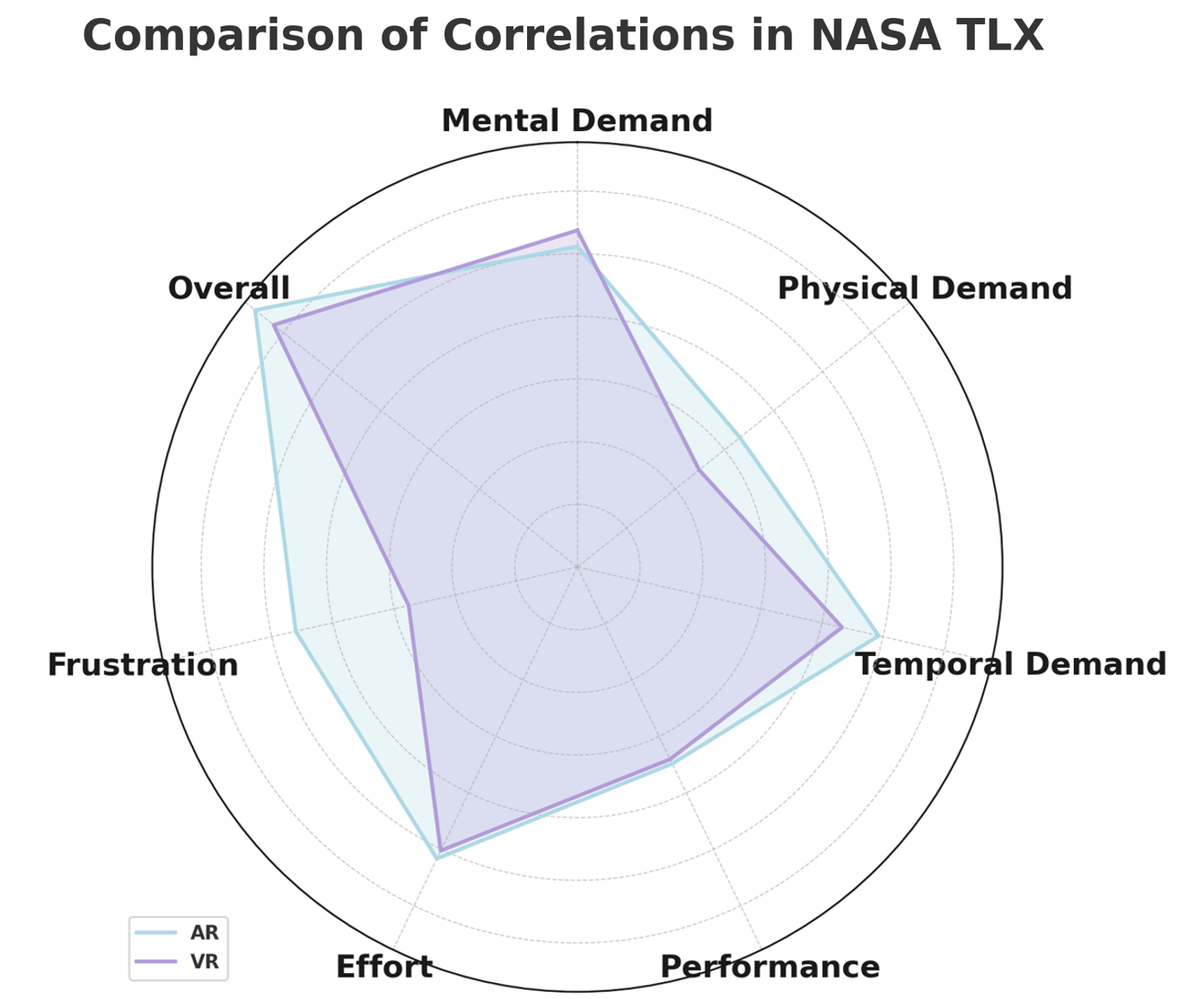}
    \caption{Correlation Between Categories in NASA TLX}
    \Description{A radar graph shows the correlation between categories in NASA TLX results from AR and VR conditions.}
    \label{fig:nasa_tlx_correlation}
  \end{minipage}
\end{figure}

\subsection{NASA TLX and Presence}
A paired samples t-tests was conducted on NASA TLX scores from each dimension with both conditions. None of the dimensions show a statistically significant difference between the AR and VR conditions (using a conventional alpha level of 0.05). As shown in ~\autoref{fig:nasa_tlx_boxplot}, the mental demand scores are generally higher in VR than in AR, and frustration in both conditions is comparable. In addition, correlation analysis (refer to Figure~\ref{fig:nasa_tlx_correlation}) revealed that a smaller radar area for frustration in VR, which indicated frustration is less interconnected with other workload aspects in VR than in AR. These findings suggest that users might be more tolerant of errors or learning curves associated with a fully immersive experience.

\begin{figure}
  \begin{minipage}[t]{1\linewidth}
    \centering
    \includegraphics[width=1\textwidth]{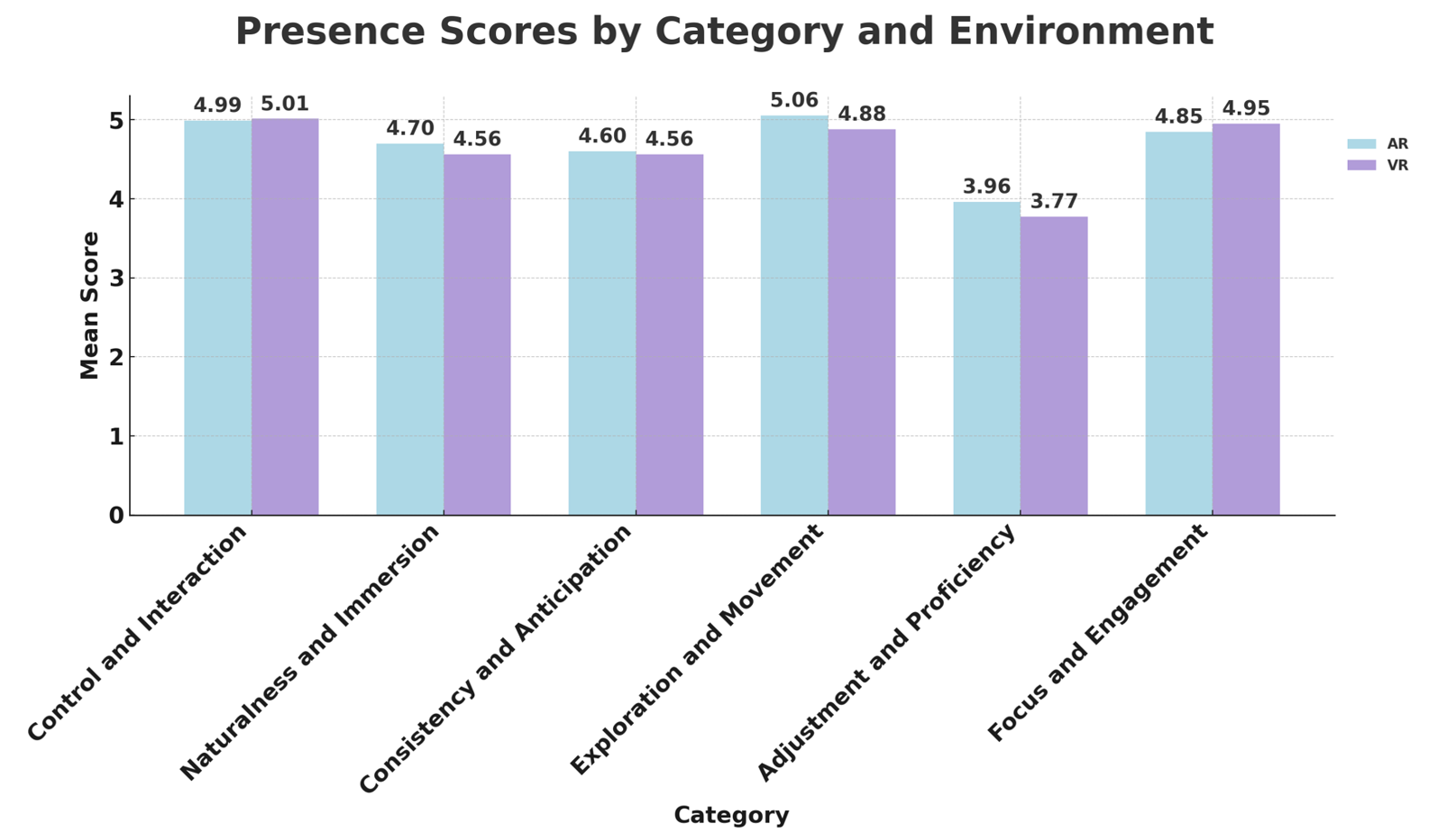}
    \caption{Comparison of Presence Scores}
    \Description{A bar chart shows that presence scores from AR and VR environments.}
    \label{fig:presence}
  \end{minipage}
  \end{figure}

The overall average presence score across all questions and environments is 4.56 on a scale of 1 to 7, suggesting a moderate to high level of presence experienced by participants. Due to the unequal variances across conditions, we conducted Wilcoxon signed-rank tests on aggregated categories of questionnaire items to compare the presence scores between AR and VR conditions. The results revealed no statistically significant differences across most categories at a alpha level of 0.05, suggesting comparable immersion and interaction for participants in both settings. In ~\autoref{fig:presence}, However, some categories approached significance, suggesting nuanced differences in participant experiences between the two environments. For example, categories of naturalness and immersion approached significance (p = 0.051), indicating a possible difference in how natural and immersive the environments were perceived. Also, categories of adjustment and proficiency approached significance (p = 0.078), hinting at potential differences in how participants adjusted to and interacted with the environments.

\subsection{Video Analysis and Interview Feedback}

\begin{figure}
  \begin{minipage}[t]{1\linewidth}
    \centering
    \includegraphics[width=1\textwidth]{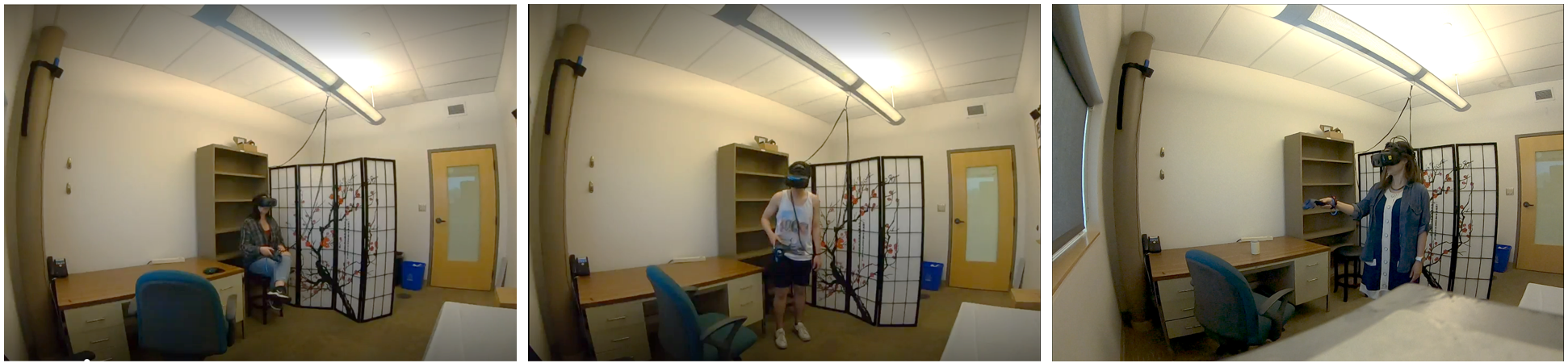}
    \caption{Postures in the Experiment}
    \Description{A photo shows main postures in the experiment.}
    \label{fig:mainPostures}
  \end{minipage}%
  \end{figure}

\begin{figure}
  \begin{minipage}[t]{0.9\linewidth}
    \centering
    \includegraphics[width=1\textwidth]{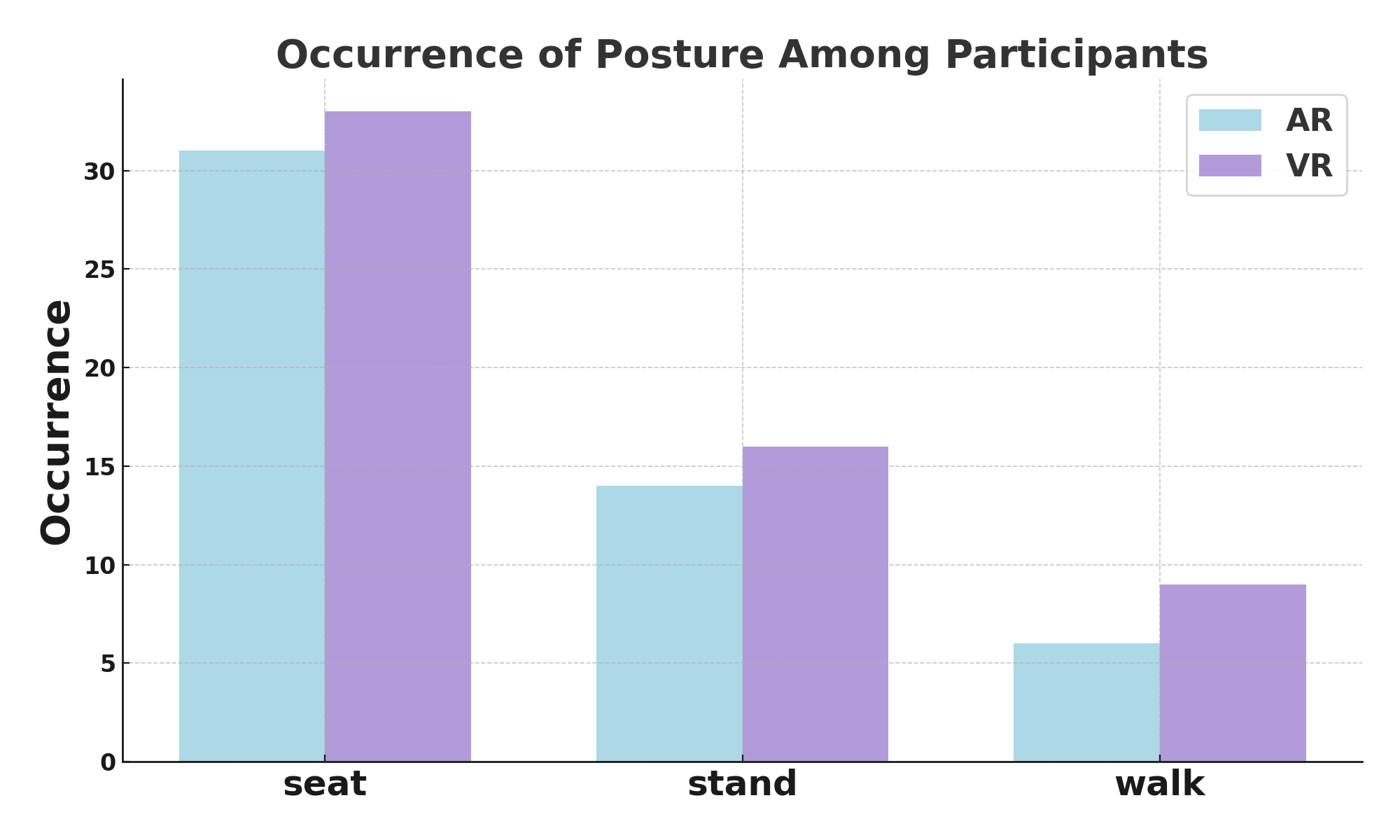}
    \caption{Participants Postures Distribution in AR and VR}
    \Description{A bar chart shows the distribution of participants' postures, including sitting, standing, and walking in AR and VR}
    \label{fig:postures}
  \end{minipage}
\end{figure}

In their study, Lisle et al. found that the tether of the HMD caused concern among participants, potentially affecting their choice to walk in the VR environment ~\cite{LIS+23}. To address this issue, a hook was used in our study to hold the tether above the participants, eliminating distraction and preventing incidents. After collecting postures shown by each participant in each environment, three major postures were identified: sitting, standing, and walking. A chi-square test revealed no significant difference between conditions across three postures that we found among participants throughout the entire study (~\autoref{fig:mainPostures}). Additionally, participants showed more varied postures in VR compared to AR; particularly an increased amount of walking activity. This result contradicts our hypothesis H3 as it shows more physical navigation occurring in VR than AR. Furthermore, 22 participants demonstrated greater variation within similar postures in VR for sitting and standing compared to AR (~\autoref{fig:postures}). 

Interestingly, six participants replied that they did not really notice that they were in different environments or the differences between AR and VR. \textit{``I did not notice that I am in different environments, all I noticed is that oh I am more familiar with how to use these tools.''} Nevertheless, for participants who noticed the differences, they indicated that they have greater situational awareness in AR, enabling them to better understand the space they are navigating and facilitating the sensemaking process. Video analysis further supports this, showing that five participants walked a longer distance in AR compared to VR while exploring visualizations under both conditions.

Similar to Whitlock et al. discovered in their study ~\cite{WHI+20}, three participants mentioned that text-reading in AR is difficult due to the presence of physical objects in the background. Additionally, two participants brought up concerns about clutter and its distracting nature. However, as shown in ~\autoref{fig:teaser}, there does not appear to be any significant clutter in the room - it resembles a typical office setting. This observation validates that ordinary objects in an AR environment can disrupt users when they are engaged in demanding cognitive tasks, even without influencing performance ~\cite{SAT+21}. Nonetheless, participants prefer AR due to the safeness and also the natural feeling it provides. \textit{``VR feels fake, I like AR better, cause VR just a little bit silly.''}, \textit{``AR is really cool. You safely see everything around you, but at the same time, you are totally doing some virtual reality thing...''}, \textit{``In AR, can see the real world, feels more natural.''}, \textit{``I kinda like the AR more, because it feels more like I am here.''}

Half of the participants mentioned that practice and training helped them become familiar with the tools and data, allowing them to answer questions more quickly. \textit{``You just need to get used to using it and focus on it.''}, \textit{``As long as you're accustomed to the tool, answering the question is easy.''} Individual preferences also play a role in how participants choose to navigate through the data: \textit{``Once I figured out how to manipulate the visualization for my needs, it became much easier.''}, \textit{``Being able to move around was really helpful...''}.

\section{Discussion}
The NASA TLX responses gave us insights into the importance of immersion and engagement in virtual environments for enhancing the user experience. The high correlation between Mental Demand and Effort, along with the low correlation between Frustration and Effort in VR, suggests that even when tasks in VR require significant effort and cognitive load, the immersive experience may help alleviate feelings of frustration. The immersive nature of VR could potentially induce a \emph{flow} state where participants become so engrossed in the activity that they lose track of external frustrations. Additionally, by offering a more seamless and entirely digital environment, users are spared from having to integrate digital elements into the physical world, which can reduce potential sources of frustration. In designing future iterations of immersive analysis tools, it is important to consider the complexity of tasks and provide options for different realities based on cognitive load demands. For instance, cross-virtuality systems could offer seamless integration and transitions between different reality implementations. This would allow users involved in extensive data analysis or prolonged sensemaking processes to enter a VR environment through a ``portal'' while being able to switch back to AR when tackling simpler tasks or wanting to maintain the connection with the real world.

Based on the responses to the presence questionnaire, participants felt that both environments provided realistic experiences and allowed for seamless exploration and movement, creating a strong sense of involvement and immersion. However, technical issues served as reminders of the artificial nature of the environment and could result in frustration or disengagement. In addition to preventing technical difficulties, it is imperative for HCI researchers to optimize visual presentation and enhance input device responsiveness. One approach could be implementing a hybrid user interface~\cite{FEI+91} that incorporates familiar input methods like tablet or smartwatch interfaces and also compensates for the downsides of mixed reality HMDs, which would help ease users into the 3D interaction while enhancing their overall presence in their experience.

  \begin{figure}
  \begin{minipage}[t]{1\linewidth}
    \centering
    \includegraphics[width=1\textwidth]{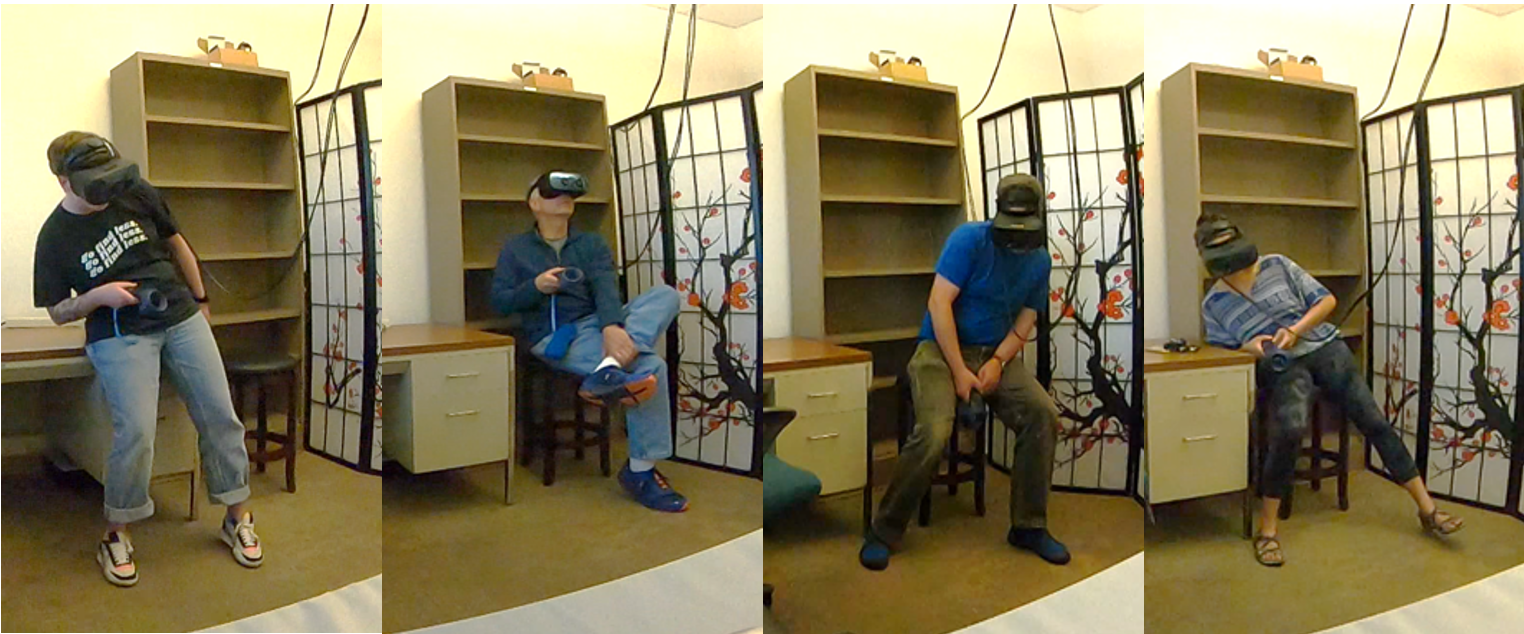}
    \caption{Various of Posture in VR}
    \Description{A combined photo showed various of postures in VR.}
    \label{fig:variation}
  \end{minipage}
\end{figure}

The posture analysis of the participants indicates that they exhibited a wider range of postures in VR compared to AR, with notably more walking in the VR environment. This suggests that participants felt secure and fully engaged with the tasks within the immersive virtual environment. Further detailed findings also affirm that immersion enhances engagement and focus. For instance, participants displayed a greater variety of still postures, as demonstrated in ~\autoref{fig:variation}. This reflects their relaxed state while comfortably adapting their posture, indicative of being ``in the zone.'' These insights highlight to researchers that when aiming to introduce new interaction techniques or demanding tasks requiring high user concentration, increasing immersion during the design phase should be carefully considered.

The diversity of our participants suggests that our results are generalizable for users from various backgrounds. It also resulted in various perspectives from different ages and individual preferences. One participant highlighted the importance of considering \textit{``different ages of users''} and the \textit{``generation gap,''} while praising the system. Additionally, one participant expressed concerns about \textit{``looking ridiculous''} with stretching arms for interaction but became more engaged in the experiment after realizing that \textit{``no one is watching.''} This confirmed our assumption that users exhibit heightened self-awareness when they know the experimenter is observing from a distance. As anticipated, the folding screen used in this study effectively ensured privacy, indicating that future study designs should replicate real user scenarios, such as simulating personal offices to encourage natural interaction and increase participants' engagement.

\section{Limitation and Future Work}

Technical difficulties arising from unstable headset cable connectivity have caused some participants to experience issues during the experiments, potentially introducing bias to the system's usability evaluation and overall user experience. While we were able to address these hardware issues as they occurred, future endeavors must take into account the potential for technical challenges during studies and establish contingency plans to mitigate their impact.

Our findings confirm that implementation with AR or VR is influenced by the immersive environment's nature, required efforts, and individual preferences. Our upcoming work will concentrate on integrating various realities into the immersive analytics platform, affording users the option to select their preferred reality for specific tasks or at any given time. Additionally, this involves incorporating input devices and interfaces with a minimal learning curve so that users can reduce cognitive load while becoming familiar with analysis tools and 3D visualization.

Although this study has been conducted with the best mixed reality device available on the market, the study results will vary with different environments and headsets. We are planning to validate the results with other mixed reality HMDs and environment setups.

\section{Conclusion}

This study compares user sensemaking strategies with immersive analytics using quantitative data in both AR and VR environments. By simulating the physical environment in VR and using a mixed-reality headset for both AR and VR conditions, we significantly mitigate confounding variables. Our results show comparable user performance in both conditions, as well as similar subjective feedback regarding presence experience, overall workload, and system usability. However, we discovered that users exhibit a higher tolerance for mental demand and effort in VR than in AR. Regarding navigation, more walking was shown in VR while further travel distance was found in AR. Additionally, users showed more posture variants in VR, indicating greater engagement and focus on tasks at hand. This work provides insights into how different realities facilitate the sensemaking process and offers guidelines for future immersive analytics platform design.

\begin{acks}
This work was supported by the National Science Foundation (NSF) awards
2327569, 2238313, 2223432, 2223459, 2106590, 2016714, 2037417. This work was also supported by the Defense Advanced Research Projects Agency (DARPA) award HR00112110011 and the Office of Naval Research (ONR) award N00014-21-1-2949.
\end{acks}

\bibliographystyle{ACM-Reference-Format}
\bibliography{LBW}

\appendix

\end{document}